\documentclass[prl,reprint,amsmath,amssymb,floatfix,citeautoscript,superscriptaddress]{revtex4-1}
\pdfoutput=1
\usepackage{cancel}
\usepackage{amsmath}
\usepackage{physics}
\usepackage{graphicx}
\usepackage{amssymb}
\usepackage{amsthm}
\usepackage{bm}
\usepackage{dcolumn}
\usepackage{braket}
\usepackage{longtable}
\usepackage{lipsum}

\usepackage{ragged2e}
\usepackage{txfonts}

\usepackage{color}
\usepackage[usenames,dvipsnames]{xcolor}
\definecolor{myblue}{rgb}{0,0,1}
\usepackage[breaklinks=true,colorlinks=true,linkcolor=myblue,urlcolor=myblue,citecolor=myblue]{hyperref}

\newcommand{\vx}{{\bm{x}}}

\allowdisplaybreaks

\begin{document}

\title{Full-Frequency GW without Frequency}

\author{Sylvia J. Bintrim}
\affiliation{Department of Chemistry,
Columbia University, New York, New York 10027, USA}
\author{Timothy C. Berkelbach}
\email{tim.berkelbach@gmail.com}
\affiliation{Department of Chemistry,
Columbia University, New York, New York 10027, USA}
\affiliation{Center for Computational Quantum Physics, Flatiron Institute, New York, New York 10010, USA}

\begin{abstract}
Efficient computer implementations of the GW approximation must approximate a
numerically challenging frequency integral; the integral can be performed
analytically, but doing so leads to an expensive implementation whose
computational cost scales as $O(N^6)$ where $N$ is the size of the system.  Here
we introduce a new formulation of the full-frequency GW approximation by exactly
recasting it as an eigenvalue problem in an expanded space.  This new
formulation (1) avoids the use of time or frequency grids, (2) naturally
precludes the common ``diagonal'' approximation, (3) enables common iterative
eigensolvers that reduce the canonical scaling to $O(N^5)$, and (4) enables a
density-fitted implementation that reduces the scaling to $O(N^4)$.  We
numerically verify these scaling behaviors and test a variety of approximations
that are motivated by this new formulation.  In this new formulation, the
relation of the GW approximation to configuration interaction, coupled-cluster
theory, and the algebraic diagrammatic construction is made especially apparent,
providing a new direction for improvements to the GW approximation. 
\end{abstract}

\maketitle

Green's function approaches based on time-dependent many-body perturbation
theory provide an economical description of excitation energies and spectral
intensities.  For the one-particle Green's function, which describes electron
addition and removal processes, the GW approximation to the
self-energy~\cite{Hedin1965} performs well for weakly correlated insulators and
metals~\cite{Hybertsen1986}, which has partially motivated its application to
molecules~\cite{Tiago2006,Bruneval2012,VanSetten2013,VanSetten2015} (here and
throughout we are considering the common non-self-consistent G$_0$W$_0$
approximation, which we call the GW approximation for simplicity).  The size of
systems that can be studied with the GW approximation is determined by the
implementation, which can be characterized by its asymptotic scaling with the
system size $N$, ranging from $O(N^3)$ to $O(N^6)$ with widely varying
prefactors~\cite{Foerster2011,Golze2019}.  

GW implementations can be distinguished 
based on their handling of a numerically challenging frequency integral, which is relatively uncommon in
quantum chemical methods. 
The earliest works used a generalized plasmon pole model to approximate the dielectric function
and thus integrate analytically~\cite{Hybertsen1986,Godby1989,Larson2013}.  More
sophisticated approaches treat the full frequency dependence using numerical
integration techniques such as analytic
continuation~\cite{Rieger1999,Giustino2010,Ren2012,Wilhelm2016,Golze2018,Zhu2020} and 
contour deformation~\cite{Godby1988,Lebegue2003,Govoni2015,Golze2018,Zhu2020}.
These latter methods introduce numerical errors, but ones that in principle can
be eliminated with increasing cost (e.g.~the frequency integration grid or the
fitting of the self-energy on the imaginary frequency axis).  The final class of
methods are numerically exact within a given single-particle basis set and require 
the explicit enumeration of
all neutral excitations
energies~\cite{Hedin1991,Tiago2006,Bruneval2012,VanSetten2013,Bruneval2016}, typically
calculated within the random-phase approximation (RPA).  This explicit enumeration,
i.e.~a sum over states, dominates the cost of such a GW calculation due to its
$O(N^6)$ scaling.  This exact handling of the full frequency dependence is the
type that we address in the present work.
We note that this class of methods still constructs a frequency-dependent self-energy, 
which is used to solve the quasiparticle equation for each excitation.  In this work, we present a
new formulation of the GW approximation by recasting it as an eigenvalue problem in an expanded
space, and \textit{a frequency variable never appears}.

Within Green's function theories, charged excitation energies $E_n$, i.e.~ionization potentials (IPs)
and electron affinities (EAs), are found as the poles of the
one-particle Green's function matrix $\mathbf{G}(\omega)$ via the eigenvalue problem
\begin{equation}
\label{eq:g_eig}
\left[\mathbf{G}^{-1}(\omega=E_n)\right] \bm{R}_n 
    = \left[\mathbf{f} + \mathbf{\Sigma}_\mathrm{c}(\omega=E_n)\right]
        \bm{R}_n
    = E_n \bm{R}_n,
\end{equation}
where $\mathbf{f} = \mathbf{h} + \mathbf{J} + \mathbf{K}$ is the Fock matrix,
$\mathbf{h}$ is the kinetic and external potential energy matrix, $\mathbf{J}$ is
the Hartree matrix, $\mathbf{K}$ is the exchange matrix, and
$\mathbf{\Sigma}_\mathrm{c}(\omega)$ is the correlation part of the self-energy
matrix.
In practice, we typically work in a basis of orbitals $\phi_p(\vx)$ that diagonalize a mean-field
Green's function, which serves as the reference and defines the orbital energies $\varepsilon_p$. 
As usual, the $O$ occupied orbitals will be indexed
by $i,j,k,l$, the $V$ unoccupied orbitals by $a,b,c,d$, and generic orbitals by $p,q,r,s$.
For simplicity, we will assume real orbitals.
In this basis, we have $\mathbf{f} = \mathbf{\varepsilon} + \mathbf{K} - \mathbf{V}_\mathrm{xc}$
where $\mathbf{V}_\mathrm{xc}$ is the exchange-correlation potential matrix.
Note that for a HF reference, $\mathbf{K}-\mathbf{V}_{\mathrm{xc}} = 0$.

In the GW approximation, the self-energy is calculated to lowest-order in the
screened Coulomb interaction $W$, which gives rise to the aforementioned frequency integral,
$\Sigma_\mathrm{c}(\omega) = (i/2\pi) \int d\omega^\prime e^{i\eta \omega^\prime}
    G(\omega+\omega^\prime) W_\mathrm{p}(\omega^\prime)$
where $W_\mathrm{p} = W - v$ is the polarized part of the screened Coulomb interaction.
When the polarizability that enters $W_\mathrm{p}$ is expressible by a spectral representation,
then the frequency integration can be performed analytically to 
yield~\cite{Hedin1991,Tiago2006,Bruneval2012,VanSetten2013,Bruneval2016}
\begin{equation}
\label{eq:gw_sigma}
\begin{split}
\left[\Sigma_\mathrm{c}(\omega)\right]_{pq} &= \sum_\nu \left\{ 
    \sum_j \left[ \frac{M_{pj}^{\nu} M_{qj}^{\nu}}
                       {\omega - (\varepsilon_j - \Omega_\nu) - i\eta} \right] \right. \\
    &\hspace{4em} \left. +\sum_b \left[ \frac{M_{bp}^{\nu} M_{bq}^{\nu}}
                        {\omega - (\varepsilon_b + \Omega_\nu) + i\eta} \right]
             \right\}.
\end{split}
\end{equation}
where
$M_{pq}^{\nu} = \int d\vx_1 d\vx_2 \phi_p(\vx_1) \phi_q (\vx_1) r_{12}^{-1} \rho_\nu(\vx_2)$,
$\Omega_\nu$ are neutral excitation energies, and $\rho_\nu(\vx)$ are transition densities.

Although any theory of neutral excitations can be used to calculate the polarizability~\cite{Lewis2019},
here we consider the Tamm-Dancoff approximation (TDA) to the (direct) RPA. 
Within the TDA, the neutral excitation energies and
transition density moments are defined by $\mathbf{A} \mathbf{X}^\nu = \Omega_\nu \mathbf{X}^\nu$, where
\begin{equation}
A_{ia,jb} = (\varepsilon_a - \varepsilon_i) \delta_{ab}\delta_{ij} + \langle ib|aj\rangle,
\end{equation}
$\langle pq|rs\rangle = \int d\vx_1 d\vx_2 \phi_p(\vx_1) \phi_q(\vx_2) r_{12}^{-1} \phi_r(\vx_1)\phi_s(\vx_2)$,
$\rho_\nu(\vx) = \sum_{ia} X_{ia}^{\nu} \phi_i(\vx) \phi_a(\vx)$, and
$M_{pq}^{\nu} = \sum_{ia} X_{ia}^{\nu} \langle pi|qa\rangle$.
Diagramatically, such a self-energy has screening due to infinite-order, 
\textit{forward time-ordered} ring (or bubble) diagrams.
The algebraic form Eq.~(\ref{eq:gw_sigma}) assumes
that \textit{all} eigenvalues and eigenvectors of the $\mathbf{A}$ matrix have been calculated, which
implies a canonical $O(N^6)$ scaling, as discussed in the introduction.

\begin{figure}[b!]
  \centering
    \includegraphics[scale=0.85]{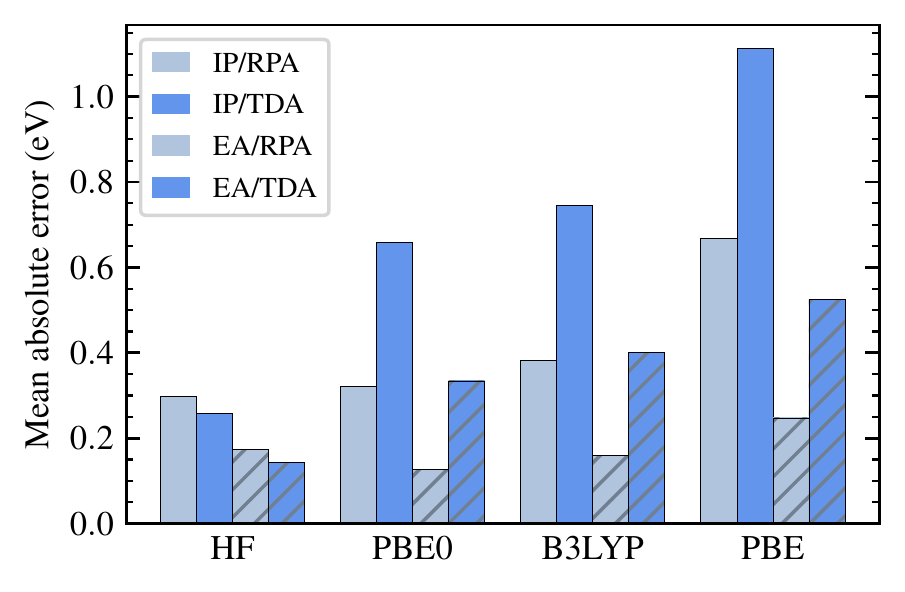}
  \caption{
Mean absolute errors in the first IP and first EA calculated using GW with TDA and RPA screening for 
various mean-field references. 
Using the $O(N^6)$ implementation in PySCF~\cite{Sun2018,Sun2020}, we performed calculations
on the smallest 91 molecules in the GW100 test set~\cite{VanSetten2015} in the def2-TZVPP 
basis~\cite{Weigend2005}.
Error is calculated with respect to 
to $\Delta$CCSD(T) for the IP~\cite{Krause2015} and EOM-CCSD for the
EA~\cite{Lange2018}.
}
  \label{fig:tda_rpa}
\end{figure}

In the GW community, RPA screening is much more commonly implemented without the TDA.  
Although the frequency-free implementation of the GW
approximation that we present here is far simpler to formulate within the TDA,
many of the same ideas can be applied for the case of RPA screening, which we discuss
in the Supplemental Information.  In particular, we show that a frequency-free formulation with RPA
screening exists; however, it is less conducive to cost reductions.
Moreover, in Fig.~\ref{fig:tda_rpa}, we show that results obtained with TDA screening are of similar
accuracy to those obtained with RPA screening, especially when based on a HF reference, empirically
justifying our focus on TDA screening.

In order to make progress on a frequency-free implementation that avoids the
explicit sum over states in Eq.~(\ref{eq:gw_sigma}), we define a vector space of
excitations corresponding to one hole (1h), one particle (1p), two holes and one
particle (2h1p), and two particles and one hole (2p1h).  A vector in this space
has elements $\bm{R} = (r_i, r_a, r_{i[jb]}, r_{[jb]a})$.
The notation of the 2h1p and 2p1h amplitudes indicates that the $j\rightarrow b$
excitation is independent of the other particle or hole index, i.e.~the amplitudes
do not obey any antisymmetry as they do in determinantal approaches.
We define a frequency-independent ``super-matrix'' $\mathbf{H}$,
\begin{equation}
\mathbf{H} =
\left(
\begin{array}{ccc}
\mathbf{f} & \mathbf{V}^\mathrm{2h1p} & \mathbf{V}^\mathrm{2p1h} \\
(\mathbf{V}^\mathrm{2h1p})^\dagger & \mathbf{C}^\mathrm{2h1p} & \mathbf{0} \\
(\mathbf{V}^\mathrm{2p1h})^\dagger & \mathbf{0} & \mathbf{C}^\mathrm{2p1h} 
\end{array}
\right),
\end{equation}
where 
$\mathbf{C}^{\mathrm{2h1p}} = \bm{\varepsilon}^{\mathrm{1h}}\oplus (-\mathbf{A})$ and
$\mathbf{C}^{\mathrm{2p1h}} = \bm{\varepsilon}^{\mathrm{1p}}\oplus \mathbf{A}$
with matrix elements
\begin{subequations}
\begin{align}
V^\mathrm{2h1p}_{p,k[lc]} &= \langle pc|kl\rangle \\
V^\mathrm{2p1h}_{p,[kc]d} &= \langle pk|dc\rangle \\
C^\mathrm{2h1p}_{i[ja],k[lc]} &=  \left[(\varepsilon_i+\varepsilon_j-\varepsilon_a)\delta_{jl}\delta_{ac} 
    - \langle jc|al\rangle\right]\delta_{ik}\\
C^\mathrm{2p1h}_{[ia]b,[kc]d} &=  \left[(\varepsilon_a+\varepsilon_b-\varepsilon_i)\delta_{ik}\delta_{ac} 
    + \langle ak|ic\rangle \right]\delta_{bd}.
\end{align}
\end{subequations}
This super-matrix can be downfolded into the 1h+1p space, leading to a frequency-dependent eigenvalue
problem of the form Eq.~(\ref{eq:g_eig}), with
\begin{equation}
\begin{split}
\mathbf{\Sigma}(\omega) 
    &= \mathbf{V}^{\mathrm{2h1p}} 
    \left[\omega\mathbf{1}-\mathbf{C}^{\mathrm{2h1p}}\right]^{-1}
    [\mathbf{V}^{\mathrm{2h1p}}]^\dagger \\
    &\hspace{1em} + \mathbf{V}^{\mathrm{2p1h}}
    \left[\omega\mathbf{1}-\mathbf{C}^{\mathrm{2p1h}}\right]^{-1} 
    [\mathbf{V}^{\mathrm{2p1h}}]^\dagger.
\end{split}
\end{equation}
It is straightforward to check that this frequency-dependent matrix, arising
from the downfolding of the 2h1p and 2p1h spaces, is precisely the correlation
part of the GW self-energy. 
The above presentation closely follows the algebraic diagrammatic construction (ADC)
literature~\cite{Schirmer1983,VonNiessen1984}.  In particular, the above
theory,~i.e. the GW approximation with TDA screening, is a strict but severe
approximation to the ADC(3) and 2p1h-TDA methods~\cite{Schirmer1978}.  Diagramatically, the latter
two theories include many vertex corrections beyond the GW approximation,
including ladder and exchange diagrams.  An analogous approach was also used recently to
formulate an efficient renormalized second-order Green's function theory~\cite{Backhouse2020}
and similar conceptual ideas were discussed in the context of double excitations in time-dependent
density functional theory~\cite{Romaniello2009}.

Importantly, the frequency-independent super-matrix form of the GW approximation enables
the use of iterative eigensolvers that lower the computational scaling.
Matrix-vector multiplication is given by $\mathbf{H}\bm{R} = \bm{\sigma}$, with 
\begin{subequations}
\begin{align}
\begin{split}
\sigma_i &= \sum_j f_{ij} r_j + \sum_b f_{ib} r_b \\
    &\hspace{1em} + \sum_{klc} \langle ic|kl\rangle r_{k[lc]} + \sum_{kcd} \langle ik|dc\rangle r_{[kc]d}
\end{split} \\
\begin{split}
\sigma_a &= \sum_j f_{aj} r_j + \sum_b f_{ab} r_b \\
    &\hspace{1em} + \sum_{klc} \langle ac|kl\rangle r_{k[lc]} + \sum_{kcd} \langle ak|dc\rangle r_{[kc]d}
\end{split} \\
\begin{split}
\sigma_{i[ja]} &= \sum_k \langle ka|ij\rangle r_k + \sum_b \langle ba|ij\rangle r_b \\
    &\hspace{1em} + (\varepsilon_i+\varepsilon_j-\varepsilon_a) r_{i[ja]}
        - \sum_{lc} \langle jc|al\rangle r_{i[lc]} 
\end{split} \\
\begin{split}
\sigma_{[ia]b} &= \sum_j \langle ji|ba\rangle r_j + \sum_c \langle ci|ba\rangle r_c \\
    &\hspace{1em} + (\varepsilon_a+\varepsilon_b-\varepsilon_i) r_{[ia]b} 
        + \sum_{kc} \langle ak|ic\rangle r_{[kc]b},
\end{split}
\end{align}
\end{subequations}
where all indices correspond to spin-orbitals.  For a restricted, closed-shell reference,
spin-free equations are straightforward to derive and are given in the Supplemental Information. 
Clearly, the above equations have no worse than $O(N^5)$ scaling (specifically
$O^2V^3$ for moderately sized basis sets), which is a significant improvement over the
$O(N^6)$ scaling exhibited by the sum-over-states implementation.
Furthermore, because only Coulomb-type electron repulsion integrals 
are used in the direct TDA (or RPA), the scaling of the 
most expensive contractions can be easily reduced by density-fitting.  For example,
if the ERIs are approximated as $(pq|rs) \approx \sum_Q B_{pq}^{Q} B_{rs}^Q$
then the worst-scaling $O(N^5)$ term can be calculated by
\begin{equation}
\sigma_{[ia]b} = 
    \sum_{Q} \sum_{kc} B_{ai}^Q B_{kc}^Q r_{[kc]b},
\end{equation}
which has two steps that scale as $O(N_\mathrm{aux} OV^2)$ or $O(N^4)$.

Although we will not show results here, we briefly describe how spectral quantites can also be
obtained iteratively with identical scalings. 
Using a spectral resolution of $\mathbf{H}$, the full Green's function is given by
$G_{pq}(\omega) = \sum_n (r_p^n r_q^n)/(\omega - E_n)$,
i.e.~the quasiparticle weight is given simply in terms of the 1p+1h elements of
the solution vector $\bm{R}$; this formulation naturally precludes the common diagonal
approximation $\Sigma_{pq}(\omega) \approx \delta_{pq} \Sigma_{pp}(\omega)$.
The matrix $\mathbf{H}$ can also be used iteratively (without diagonalization) 
to calculate the frequency-dependent self-energy
$\mathbf{\Sigma}(\omega) = \mathbf{V}^\mathrm{2h1p} \mathbf{Z}^\mathrm{2h1p}(\omega)
    + \mathbf{V}^\mathrm{2p1h} \mathbf{Z}^\mathrm{2p1h}(\omega)$
where $\mathbf{Z}(\omega)$ is a matrix that solves the linear systems of equations,
e.g.
$\left[\omega\mathbf{1}^\mathrm{2h1p} - \mathbf{C}^\mathrm{2h1p}\right] \mathbf{Z}^\mathrm{2p1h}(\omega)
    = [\mathbf{V}^\mathrm{2h1p}]^\dagger$,
which can be solved with iterative methods such as conjugate gradient or the generalized minimum
residual method.
Similarly, the Green's function can be calculated as
$\mathbf{G}(\omega) = \mathbf{P} \mathbf{Z}(\omega)$
where $\mathbf{Z}(\omega)$ solves
$\left[\omega\mathbf{1} - \mathbf{H}\right] \mathbf{Z}(\omega)
    = \mathbf{P}^\dagger$
and $\mathbf{P}$ is a matrix that projects onto the 1p+1h space.

\begin{figure}[t!]
  \centering
    \includegraphics[scale=0.9]{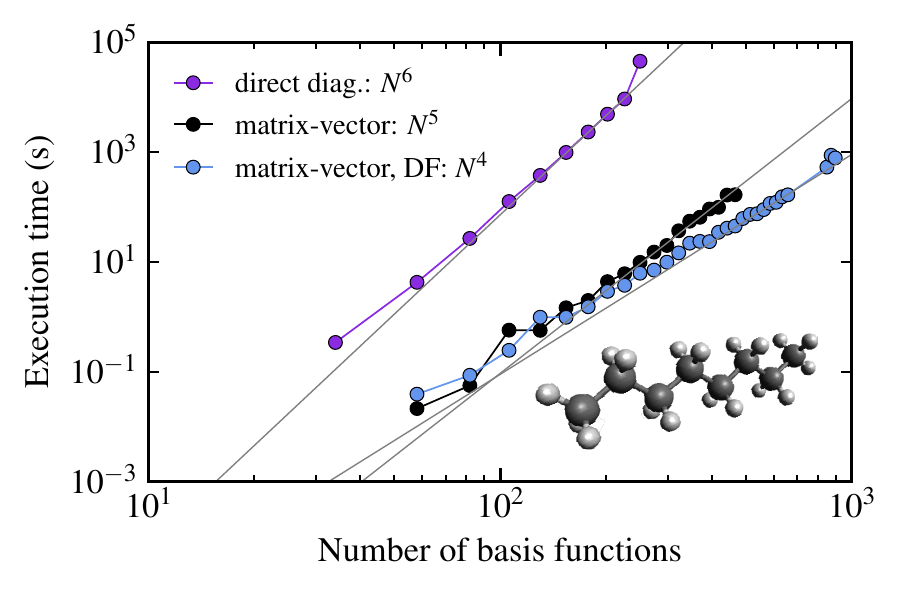}
  \caption{Timings of the sum-over-states (direct diag.), frequency-free
(matvec), and density-fitted frequency-free (matvec, DF) implementations of the
GW approximation for a series of linear alkanes in the def2-SVP basis, up to
C$_{37}$H$_{76}$.
For the sum-over-states implementation, we timed the full diagonalization of the TDA matrix;
for the frequency-free implementations, we timed the 6-8 matrix-vector multiplications required for
convergence of the Davidson algorithm.
}
  \label{fig:scaling}
\end{figure}

We have implemented the GW techniques described above in the PySCF software
package~\cite{Sun2018,Sun2020}.  To compare their costs and verify their asymptotic
scaling, we have calculated the first IP of a series of linear alkanes in
the def2-SVP basis~\cite{Weigend2005} up to C$_{37}$H$_{76}$, which has 898 basis functions 
The
execution timings of the $O(N^6)$ sum-over-states, $O(N^5)$ frequency-free, and
$O(N^4)$ density-fitted frequency-free implementations are shown in
Fig.~\ref{fig:scaling}; all calculations were performed on a single core of an
Intel Xeon Gold 6126 2.6~GHz (Skylake) CPU and density-fitted calculations used
the def2-SVP-JKFIT auxiliary basis set~\cite{Weigend2008}.
As can be seen, all methods exhibit the expected asymptotic
scaling.  Comparing the absolute execution times of the sum-over-states and
density-fitted frequency-free implementations, we obtained a speed-up of four
orders of magnitude for the C$_{10}$H$_{22}$ calculation.  For our largest
system with almost one thousand basis functions, the density-fitted
implementation required only two hours on a single core, demonstrating the
immense savings available with the advances described
here.  
The use of density fitting was found to introduce a
negligible error of around $0.01$~eV.

One challenge with using iterative eigensolvers on $\mathbf{H}$ is that the eigenvalues
of typical interest (valence ionization potentials and electron affinities) are interior
eigenvalues.  Therefore, they must be found using energy-targeting methods like shift-and-invert
or ones which maximize eigenvector overlap with a given guess vector.  We have found the latter
to work well, in conjunction with Davidson
diagonalization~\cite{Davidson1975,Tackett2002}, for valence IPs and EAs.
However, two simple alternatives exist by introducing additional approximations.

In a first approach, one can make the diagonal approximation to the Green's function and the self-energy,
seeking the self-consistent solution $E_n$ of the algebraic equation
$f_{pp} + \Sigma_{pp}(\omega=E_n) = E_n$.
This can be solved by iterative diagonalization of a modified matrix $\mathbf{H}^{(p)}$
which has deleted all 1p+1h rows and columns except that of orbital $p$.
Although the principal eigenvalue of interest is still an interior eigenvalue, this approach
eliminates all other quasiparticle energies, which can facilitate energy- or overlap-targeting
procedures.
In a second approach, one can perturbatively decouple the IP and EA parts of $\mathbf{H}$,
which will make the valence IPs and EAs into extremal eigenvalues.
For example, for the calculation of IPs, we perturbatively eliminate the 1p and 2p1h subspaces based
on their lowest-order influence on the 1h subspace, and likewise for EAs.  This leads to modified
Fock matrix elements 
\begin{subequations}
\begin{align}
F_{ii} &= f_{ii} + \sum_b \frac{|f_{ib}|^2}{f_{ii}-f_{bb}}
    + \sum_{kcd} \frac{|\langle ik|dc\rangle|^2}{\varepsilon_i+\varepsilon_k-\varepsilon_c-\varepsilon_d} \\
F_{aa} &= f_{aa} + \sum_j \frac{|f_{bj}|^2}{f_{aa}-f_{jj}}
    + \sum_{klc} \frac{|\langle ac|kl\rangle|^2}{\varepsilon_a+\varepsilon_c-\varepsilon_k-\varepsilon_k}
\end{align}
\end{subequations}
which are used in place of $f_{ii}$ and $f_{aa}$ during the matrix-vector product.
This approach has the added benefit of reducing the size of the vector spaces for the IP and EA
problems, $\bm{R}^\mathrm{IP} = (r_i, r_{i[jb]})$ and $\bm{R}^\mathrm{EA} = (r_a, r_{[jb]a})$, 
and reducing the scaling of the IP matrix-vector product to be be $O(O^3V^2)$.
Finally, we point out that the valence IPs can be made into the \textit{lowest} eigenvalues by
negating the matrix, $\mathbf{H}^\mathrm{IP} \rightarrow -\mathbf{H}^\mathrm{IP}$.
With all of these changes, the calculation of IPs and EAs within the GW approximation looks
quite similar to that within the IP/EA-EOM-CCSD approximation~\cite{Stanton1994,Krylov2008}.
The decoupling of the IP and EA spaces is also common in the ADC literature, and referred to
as a non-Dyson approach~\cite{Schirmer1998}.  We will use the same language, and evaluate the performance of 
the Dyson (coupled IP and EA) and non-Dyson (perturbatively decoupled IP and EA) GW approximation.

\begin{figure}[!t]
  \centering
    \includegraphics[scale=0.8]{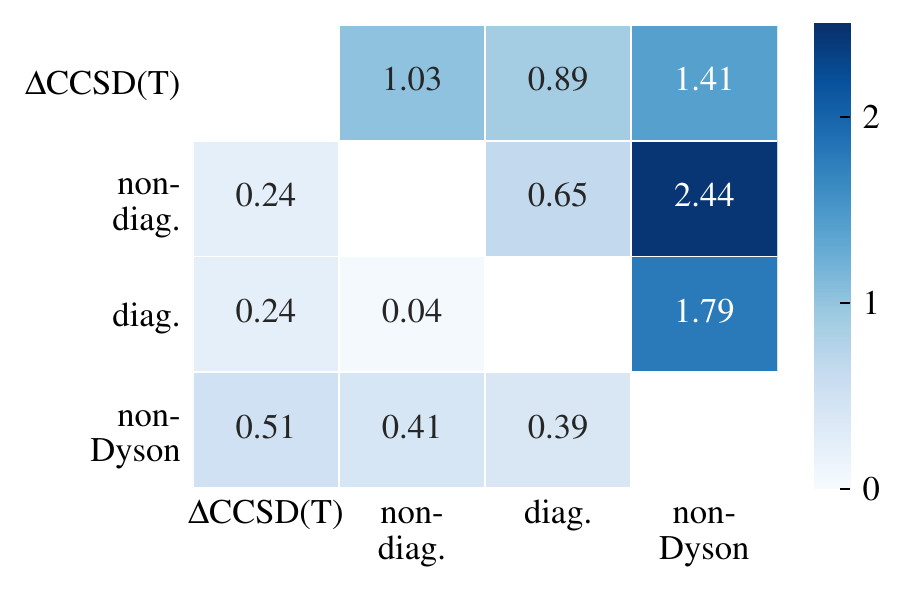}
  \caption{Impact of various approximations on the first IP of the molecules in the GW100 test set~\cite{VanSetten2015}, using a HF reference.  Mean absolute deviations are given in the lower triangle and maximum absolute deviations are given in the upper triangle (all in eV).}
  \label{fig:HF_heatmap}
\end{figure}

To assess the effect of the diagonal approximation and perturbative decoupling,
we used our frequency-free $O(N^5)$ GW implementation
to calculate the first IP of all the molecules in the GW100 test set~\cite{VanSetten2015}.
In Fig.~\ref{fig:HF_heatmap}, we compare the non-diagonal, diagonal, and non-diagonal perturbatively decoupled
(``non-Dyson'') GW results among themselves and to $\Delta$CCSD(T) results~\cite{Krause2015}, 
using a HF reference.
As shown in Fig.~\ref{fig:tda_rpa}, the GW results have a mean absolute error of 0.24~eV, with
respect to CCSD(T).
On average, the diagonal approximation has negligible
effect (less than 0.1~eV), although a maximum deviation of 0.65~eV is
observed, indicating the potential importance of off-diagonal elements of the self-energy for 
some molecules. Perturbative decoupling (the non-Dyson GW approximation)
changes the results by 0.41~eV on average (and by as much as 2.44~eV) and
increases the mean absolute error from 0.24~eV to 0.51~eV, suggesting that it is a relatively
severe approximation.
We have performed the same analysis (not shown) for the EA, as well as for a PBE
reference; all results are qualitatively similar.

To summarize, we have shown that the typical Dyson equation formulation of the GW approximation
can be exactly reformulated as a frequency-independent eigenvalue problem in an expanded space.
In addition to providing a new conceptual framework for the GW approximation and related Green's function
theories, the new formulation was used to reduce the computational scaling from $O(N^6)$ to $O(N^4)$.
Based on our preliminary results, we expect that this frequency-free formulation of the GW approximation
will be readily applicable to systems with hundreds or thousands of atoms, likely limited by the
memory needed to store three-index quantities.

We anticipate that this eigenvalue formulation of the GW approximation will lead to
new methodological developments inspired by quantum chemical methods with
similar structure. For example, we are exploring the use of partitioning schemes
to mitigate the cost of large basis
sets~\cite{Nooijen1995,Stanton1995,Lange2020}, the introduction of vertex
corrections through the algebraic diagrammatic
construction~\cite{Schirmer1983,VonNiessen1984}, the use of renormalization and compression
to perform self-consistent GW~\cite{Backhouse2020}, and the extension towards
strongly correlated systems with multi-reference
techniques~\cite{Sokolov2018,Chatterjee2019}.  Finally, the ideas presented in
this work and these latter extensions can be applied to treat the frequency
dependence of the Bethe-Salpeter equation for neutral
excitations~\cite{Grossman2001,Bruneval2015}.

\vspace{1em}

This work was supported in part by the National Science Foundation Graduate
Research Fellowship under Grant No.~DGE-1644869 (S.J.B.) and
by the National Science Foundation under Grant No.~CHE-1848369 (T.C.B.).  We
acknowledge computing resources from Columbia University’s Shared Research
Computing Facility project, which is supported by NIH Research Facility
Improvement Grant 1G20RR030893-01, and associated funds from the New York State
Empire State Development, Division of Science Technology and Innovation (NYSTAR)
Contract C090171, both awarded April 15, 2010.
The Flatiron Institute is a division of the Simons Foundation.

\bibliography{gw}

\end{document}


\title{Supplemental Information: Full-Frequency GW without Frequency}

\author{Sylvia J. Bintrim}
\affiliation{Department of Chemistry,
Columbia University, New York, New York 10027, USA}
\author{Timothy C. Berkelbach}
\email{tim.berkelbach@gmail.com}
\affiliation{Department of Chemistry,
Columbia University, New York, New York 10027, USA}
\affiliation{Center for Computational Quantum Physics, Flatiron Institute, New   York, New York 10010, USA}

\maketitle

\renewcommand\theequation{S\arabic{equation}}

\section{Spin-free equations}
\label{app:spinfree}

For a restricted, closed-shell reference system, the spin-free matrix-vector equations for the
GW approximation with TDA screening are 
\begin{subequations}
\begin{align}
\begin{split}
\sigma_i &= \sum_j f_{ij} r_j + \sum_b f_{ib} r_b \\
    &\hspace{1em} + 2\sum_{klc} \langle ic|kl\rangle r_{k[lc]} + 2\sum_{kcd} \langle ik|dc\rangle r_{[kc]d}
\end{split} \\
\begin{split}
\sigma_a &= \sum_j f_{aj} r_j + \sum_b f_{ab} r_b \\
    &\hspace{1em} + 2\sum_{klc} \langle ac|kl\rangle r_{k[lc]} + 2\sum_{kcd} \langle ak|dc\rangle r_{[kc]d}
\end{split} \\
\begin{split}
\sigma_{i[ja]} &= \sum_k \langle ka|ij\rangle r_k + \sum_b \langle ba|ij\rangle r_b \\
    &\hspace{1em} + (\varepsilon_i+\varepsilon_j-\varepsilon_a) r_{i[ja]}
        - 2\sum_{lc} \langle jc|al\rangle r_{i[lc]} 
\end{split} \\
\begin{split}
\sigma_{[ia]b} &= \sum_j \langle ji|ba\rangle r_j + \sum_c \langle ci|ba\rangle r_c \\
    &\hspace{1em} + (\varepsilon_a+\varepsilon_b-\varepsilon_i) r_{[ia]b} 
        + 2\sum_{kc} \langle ak|ic\rangle r_{[kc]b}.
\end{split}
\end{align}
\end{subequations}

\section{RPA screening}
\label{app:rpa}

As discussed in the main text, the GW approximation is more commonly implemented
without the Tamm-Dancoff approximation (TDA) to the random-phase approximation (RPA).
Here we present the frequency-free version of the GW approximation with RPA screening.
We show that a frequency-free formulation exists by recasting the GW approximation as
an eigenvalue problem in an expanded space; however, the use of RPA screening precludes
the cost reductions discussed in the main text.

The direct RPA leads to the generalized eigenvalue equation
\begin{equation}
\label{eq:rpa}
\mathbf{M}
\left(
\begin{array}{cc}
\mathbf{X} & \mathbf{Y} \\
\mathbf{Y} & \mathbf{X}
\end{array}
\right)
= 
\mathbf{N}
\left(
\begin{array}{cc}
\mathbf{X} & \mathbf{Y} \\
\mathbf{Y} & \mathbf{X}
\end{array}
\right)
\left(
\begin{array}{cc}
\mathbf{\Omega_+} & 0 \\
0 & -\mathbf{\Omega_+}
\end{array}
\right)
\end{equation}
where
\begin{subequations}
\begin{align}
\mathbf{M} &= \left(
    \begin{array}{cc}
    \mathbf{A}   & \mathbf{B} \\
    \mathbf{B} & \mathbf{A}
    \end{array}
\right) \\
\mathbf{N} &= \left(
    \begin{array}{cc}
    \mathbf{1} & \mathbf{0} \\
    \mathbf{0} & -\mathbf{1}
    \end{array}
\right) \\
A_{ia,jb} &= (\varepsilon_a - \varepsilon_i) \delta_{ab}\delta_{ij} + \langle ib|aj\rangle, \\ 
B_{ia,jb} &= \langle ij|ab\rangle
\end{align}
\end{subequations}
and $\mathbf{\Omega}_+$ is a diagonal matrix of positive excitation energies.
The correlation part of the GW self-energy is
\begin{equation}
\label{eq:gw_sigma_rpa}
\begin{split}
\left[\Sigma_\mathrm{c}(\omega)\right]_{pq} &= \sum_\nu^{\Omega_\nu > 0} \left\{ 
    \sum_j \left[ \frac{M_{pj}^{\nu} M_{qj}^{\nu}}
                       {\omega - (\varepsilon_j - \Omega_\nu) - i\eta} \right] \right. \\
    &\hspace{4em} \left. +\sum_b \left[ \frac{M_{bp}^{\nu} M_{bq}^{\nu}}
                        {\omega - (\varepsilon_b + \Omega_\nu) + i\eta} \right]
             \right\}.
\end{split}
\end{equation}
where
\begin{align}
\rho_\nu(\vx) &= \sum_{ai}\left( X_{ia}^\nu + Y_{ia}^\nu \right) \phi_i(\vx)\phi_a(\vx), \\
M_{pq}^{\nu}
    &= \sum_{ia} \left( X_{ia}^{\nu} + Y_{ia}^{\nu} \right) \langle pi|qa\rangle, \\
\delta_{\mu\nu} &= \sum_{ia} \left( X_{ia}^\mu X_{ia}^\nu - Y_{ia}^\mu Y_{ia}^\nu \right)
\end{align}
and we emphasize that only the positive-energy excitations are included in the
summation.

The GW approximation with RPA screening can be implemented in a manner that is similar to that
presented in the main text,
with a few changes.  First, the trial vector amplitudes must be expanded to include the deexcitation
space of the RPA matrix,
\begin{equation}
\bm{R} = (r_i, r_a, r_{i[jb]}, \bar{r}_{i[jb]}, r_{[jb]a}, \bar{r}_{[jb]a})
\end{equation}
Second, because the GW self-energy expression (\ref{eq:gw_sigma_rpa}) only includes positive 
excitation energies, we introduce
a matrix modification that eliminates the influence of the negative excitation energies,
$\mathbf{M} \rightarrow \mathbf{\tilde{M}}$ with
\begin{align}
\mathbf{\tilde{M}} = \mathbf{M} + \eta \mathbf{N} \Theta(-\mathbf{NM})
\end{align}
where $\eta$ is a large number and $\Theta$ is the step function.
This gives a modified eigenvalue equation
\begin{equation}
\label{eq:rpa_proj}
\mathbf{\tilde{M}}
\left(
\begin{array}{cc}
\mathbf{X} & \mathbf{Y} \\
\mathbf{Y} & \mathbf{X}
\end{array}
\right)
= 
\mathbf{N}
\left(
\begin{array}{cc}
\mathbf{X} & \mathbf{Y} \\
\mathbf{Y} & \mathbf{X}
\end{array}
\right)
\left(
\begin{array}{cc}
\mathbf{\Omega_+} & 0 \\
0 & \eta -\mathbf{\Omega_+}
\end{array}
\right),
\end{equation}
i.e.~the negative excitation energies are shifted to large positive values.
Again we define a super-matrix,
\begin{equation}
\mathbf{H} =
\left(
\begin{array}{ccccc}
\mathbf{f} & \mathbf{V}^\mathrm{2h1p} & \mathbf{V}^\mathrm{2h1p} 
    & \mathbf{V}^\mathrm{2p1h}& \mathbf{V}^\mathrm{2p1h} \\
(\mathbf{V}^\mathrm{2h1p})^\dagger & & & & \\ 
(\mathbf{V}^\mathrm{2h1p})^\dagger 
    & \multicolumn{2}{c}{\smash{\raisebox{.5\normalbaselineskip}{$\mathbf{C}^\mathrm{2h1p}$}}}
    & \multicolumn{2}{c}{\smash{\raisebox{.5\normalbaselineskip}{$\mathbf{0}$}}} \\
(\mathbf{V}^\mathrm{2p1h})^\dagger & & & & \\ 
(\mathbf{V}^\mathrm{2p1h})^\dagger 
    & \multicolumn{2}{c}{\smash{\raisebox{.5\normalbaselineskip}{$\mathbf{0}$}}}
    & \multicolumn{2}{c}{\smash{\raisebox{.5\normalbaselineskip}{$\mathbf{C}^\mathrm{2p1h}$}}}
\end{array}
\right),
\end{equation}
where 
$\mathbf{C}^{\mathrm{2h1p}} = \bm{\varepsilon}^{\mathrm{1h}}\oplus (-\mathbf{\tilde{M}})$ and
$\mathbf{C}^{\mathrm{2p1h}} = \bm{\varepsilon}^{\mathrm{1p}}\oplus \mathbf{\tilde{M}}$,
which requires a super-metric
\begin{equation}
\bm{\mathcal{N}} =
\left(
\begin{array}{ccc}
\mathbf{1} & \mathbf{0} & \mathbf{0} \\
\mathbf{0} & \mathbf{1} \oplus \mathbf{N} & \mathbf{0} \\
\mathbf{0} & \mathbf{0} & \mathbf{1} \oplus \mathbf{N}
\end{array}
\right).
\end{equation}
This leads to the generalized eigenvalue equation
$\mathbf{H} \mathbf{R} = \bm{\mathcal{N}} \mathbf{R} \mathbf{E}$
or the non-Hermitian eigenvalue equation 
$\bm{\mathcal{N}}\mathbf{H}\mathbf{R} = \mathbf{R} \mathbf{E}$
with
matrix-vector multiplication $\bm{\mathcal{N}}\mathbf{H}\mathbf{R} = \bm{\sigma}$ given by
\begin{subequations}
\begin{align}
\begin{split}
\sigma_i &= \sum_j f_{ij} r_j + \sum_b f_{ib} r_b 
    + \sum_{klc} \langle ic|kl\rangle r_{k[lc]} + \sum_{kcd} \langle ik|dc\rangle r_{[kc]d} \\
    &\hspace{1em} + \sum_{klc} \langle ic|kl\rangle \bar{r}_{k[lc]} 
    + \sum_{kcd} \langle ik|dc\rangle \bar{r}_{[kc]d} 
\end{split} \\
\begin{split}
\sigma_a &= \sum_j f_{aj} r_j + \sum_b f_{ab} r_b 
    + \sum_{klc} \langle ac|kl\rangle r_{k[lc]} + \sum_{kcd} \langle ak|dc\rangle r_{[kc]d} \\
    &\hspace{1em} + \sum_{klc} \langle ac|kl\rangle \bar{r}_{k[lc]}
    + \sum_{kcd} \langle ak|dc\rangle \bar{r}_{[kc]d} 
\end{split}\\
\begin{split}
\sigma_{i[ja]} &= \sum_k \langle ka|ij\rangle r_k + \sum_b \langle ba|ij\rangle r_b \\
    &\hspace{1em} + \varepsilon_i r_{i[ja]} - \sum_{kb} [\mathbf{N\tilde{M}}]^\mathrm{xx}_{ja,kb} r_{i[kb]}
    - \sum_{kb} [\mathbf{N\tilde{M}}]^\mathrm{xd}_{ja,kb} \bar{r}_{i[kb]}
\end{split} \\
\begin{split}
\bar{\sigma}_{i[ja]} &= -\sum_k \langle ka|ij\rangle r_k - \sum_b \langle ba|ij\rangle r_b \\
    &\hspace{1em} + \varepsilon_i \bar{r}_{i[ja]} - \sum_{kb} [\mathbf{N\tilde{M}}]^\mathrm{dx}_{ja,kb} r_{i[kb]}
    - \sum_{kb} [\mathbf{N\tilde{M}}]^\mathrm{dd}_{ja,kb} \bar{r}_{i[kb]}
\end{split} \\
\begin{split}
\sigma_{[ia]b} &= \sum_j \langle ji|ba\rangle r_j + \sum_c \langle ci|ba\rangle r_c \\
    &\hspace{1em} + \varepsilon_b r_{[ia]b} + \sum_{jc} [\mathbf{N\tilde{M}}]^\mathrm{xx}_{ia,jc} r_{[jc]b}
    + \sum_{jc} [\mathbf{N\tilde{M}}]^\mathrm{xd}_{ia,jc} \bar{r}_{[jc]b}
\end{split} \\
\begin{split}
\bar{\sigma}_{[ia]b} &= -\sum_j \langle ji|ba\rangle r_j - \sum_c \langle ci|ba\rangle r_c \\
    &\hspace{1em} + \varepsilon_b \bar{r}_{[ia]b} + \sum_{jc} [\mathbf{N\tilde{M}}]^\mathrm{dx}_{ia,jc} r_{[jc]b}
    + \sum_{jc} [\mathbf{N\tilde{M}}]^\mathrm{dd}_{ia,jc} \bar{r}_{[jc]b}.
\end{split}
\end{align}
\end{subequations}
In the above equations, we have broken up the RPA $\mathbf{N\tilde{M}}$ matrix into its four blocks
associated with the excitation (x) and dexcitation (d) spaces.
An exact evaluation of the above equations can be done by diagonalizing the RPA matrix in order
to calculate $\tilde{M}$; this diagonalization has $O(N^6)$ asymptotic scaling.  However, we have
implemented the above in order to verify that it yields eigenvalues equal to those of the GW approximation
with RPA screening.

In an effort to reduce the scaling, we have attempted an iterative implementation of the product of
$\Theta(-\mathbf{NM})$ with a trial vector.  In particular, we approximate this matrix-vector product
using Arnoldi iteration, with requires only the Krylov space generated by matrix-vector products
of $\mathbf{NM}$ that can each be carried out with $O(N^5)$ cost or with $O(N^4)$ cost if density fitting
is used.  Although this approach was found to work for small systems, we encountered numerical issues
on larger systems.  For example, the number of Arnoldi iterations needed to approximate the matrix-vector
product was prohibitively large and the Davidson algorithm exhibited convergence issues.
We consider an iterative, low-scaling, frequency-free implementation of the GW approximation with RPA screening
to be an interesting area for future work.